\def\refWangWang{[1]} 
\def\refQinLuo{[2]} 
\def\refcodesautomata{[3]} 
\def\refshyryu{[4]} 
\def\reflam{[5]} 
\def\refhandbook{[6]} 
\def\refUnicode{[7]} 
\def\N{N} 
\def\alb{A} 
\def\alba{a} 
\def\apr{P} 
\def\apri{\ell} 
\def\stri{r} 
\def\signt{\sigma} 
\def\scode{X} 
\def\scodee{x} 
\def\sclen{L} 
\def\sigcode{\Sigma} 
\def\probcor{p} 
\def\calb{A} 
\def\capr{P} 
\def\capri{\ell} 
\def\ccode{X} 
\def\csclen{L} 
\def\calba{a} 
\def\cprobcor{p} 
\centerline{\bf Error-detecting solid codes}
\vskip\baselineskip
\centerline{Nathan Carruth}
\par
\centerline{SUSTech}
\par
\centerline{March 19, 2026}
\vskip\baselineskip
\leftline{\bf 0. Introduction}
\par
A code is called {\it solid\/} if, roughly speaking, any correctly-transmitted codeword in an arbitrarily corrupted string of codewords can still be decoded correctly and unambiguously. So-called variable-length solid codes, in which codewords may differ in length, have been studied in recent work of Wang and Wang \refWangWang\ and Qin and Luo \refQinLuo. In this short note, we observe that a construction of variable-length solid codes in \refWangWang\ based on binary codes may be extended to arbitrary $n$-ary codes. We further prove an interesting error-detection property of a specific subfamily of these variable-length solid codes, and give a concrete application to a certain type of binary code.
\par
A standard reference on coding theory in general is \refcodesautomata; for background on solid codes we refer to, e.g., \refshyryu, \reflam, and \refhandbook.
\par\vskip\baselineskip\par
\leftline{\bf 1. Construction and error detection}
\par
Let $\alb$ be any finite set, and suppose that $\{ \apr_\apri \}_{\apri = 0}^n$ is any partition of $\alb$. For any string $s \in \alb^*$, we let $|s|$ denote the number of letters from $\alb$ in $s$, and for any integer $\stri$, $1 \leq \stri \leq |s|$, we let $s[\stri]$ denote the $\stri$th letter in $s$. We define the {\it signature\/} of $s$ to be the (unique) sequence $\signt(s) = (\apri_\stri)_{\stri = 1}^{|s|}$ of nonnegative integers such that $s[\stri] \in \apr_{\apri_\stri}$ for all $\stri$.
\par
As usual, we call a code $\scode \subset \alb^*$ {\it solid\/} if (i) no prefix of any word in $\scode$ is a suffix of any word in $\scode$, and (ii) no word in $\scode$ is a subword of any other word. We have the following extension of Lemma 1 in \refWangWang.
\par\vskip0.5\baselineskip\par\noindent
PROPOSITION 1. Let $\scode$ be a code on $\alb$, and suppose that $\signt(\scode)$ is a solid code on $\{ 0, \ldots, n \}$. Then $\scode$ is solid. Thus, for any solid code $\sigcode$ on $\{ 0, \ldots, n \}$, $\signt^{-1}(\sigcode)$ is a solid code on $\alb$.
\par
{\it Proof.} Let $\scodee$, $\scodee' \in \scode$, $\scodee = ab$, $\scodee' = a'b'$, $a$, $a'$, $b$, $b'$ strings over $A$. Then clearly $\signt(\scodee) = \signt(a)\signt(b)$, $\signt(\scodee') = \signt(a')\signt(b')$, so that by solidness of $\signt(\scode)$ we have $\signt(b) \neq \signt(a')$. Similarly, if $\scodee \neq \scodee'$ then $\signt(\scodee)$ cannot be a substring of $\signt(\scodee')$. This implies that $b \neq a'$ and that if $\scodee \neq \scodee'$ then $\scodee$ cannot be a substring of $\scodee'$, showing that $\scode$ is solid. The second statement follows immediately from the first once we establish that $\signt^{-1}(\sigcode)$ is a code over $\alb$, but this is easy: set $\scode_0 = \signt^{-1}(\sigcode)$, and suppose that $\scodee_i, \scodee'_j \in \scode_0$, $i = 1, \ldots, n$, $j = 1, \ldots, m$, and $\scodee_1 \cdots \scodee_n = \scodee'_1 \cdots \scodee'_m$; then without loss of generality $\scodee_1$ is a prefix of $\scodee'_1$, hence (as before) $\signt(\scodee_1)$ is a prefix of $\signt(\scodee'_1)$, hence $\signt(\scodee_1) = \signt(\scodee'_1)$, hence $|\scodee_1| = |\scodee'_1|$, hence $\scodee_1 = \scodee'_1$.
\par\vskip0.5\baselineskip\par
Now let $\sclen : \{ 1, \ldots, n \} \rightarrow \N$ (where $\N = \{ 1, 2, \ldots \}$) be any function. It is clear that the code on $\{ 0, \ldots, n \}$ given by
$$
\{ \ell\,0^{\sclen(\ell)}\,|\,\ell \in \{ 1, \ldots, n \} \}
$$
is solid, where $0^{\sclen(\ell)}$ denotes a string of $\sclen(\ell)$ $0$s. Thus by Proposition 1 the code
$$
\scode = \{ \alba_0 \ldots \alba_k \,|\, \alba_0 \in \apr_\apri,\,\apri \geq 1,\,k = \sclen(\apri) \}
$$
is solid. (This shows, in particular, that, considered as a code over the 256 8-bit bytes, the UTF-8 encoding of Unicode (\refUnicode, 3.9, 3.10) is a solid code, though it is clearly not solid when considered as a binary code.) We show that these codes have special error-detecting properties. For any string $S \in \scode^*$, let $|S|$ continue to denote the length of $S$ as a string {\it over\/} $\alb$, and similarly let $S[\stri]$, $1 \leq \stri \leq |S|$, and $\signt(S)$ continue to denote the $\stri$th letter of $S$ and the signature of $S$ as a string over $\alb$. Suppose now that a string $S \in \scode^*$ is transmitted over a noisy channel which independently modifies each letter of $S$ according to a probability distribution $\probcor(\alba \rightarrow \alba')$, $\alba$, $\alba' \in \alb$. Then we have the following result.
\par\vskip0.5\baselineskip\par\noindent
PROPOSITION 2. Let $S$, $S'$ be transmitted and received strings, respectively, with $S \in \scode^*$. Suppose that
$$
\probcor(\alba \rightarrow \alba') = 0 \hbox{ whenever }\signt(\alba), \signt(\alba') \geq 1,\,\signt(\alba) \neq \signt(\alba').\eqno{(1)}
$$
If $S' \in \scode^*$, then $\signt(S') = \signt(S)$. If moreover
$$
\probcor(\alba \rightarrow \alba') = 0\hbox{ whenever }\signt(\alba) = \signt(\alba'),\,\alba \neq \alba',\eqno{(2)}
$$
then $S' = S$.
\par
{\it Proof.} For simplicity, for $1 \leq \stri \leq |S| = |S'|$, define $\signt_\stri = \signt(S[\stri])$, $\signt'_\stri = \signt(S'[\stri])$. Suppose that (1) holds. Now since $S, S' \in \scode^*$, we must have $\signt_1, \signt'_1 \geq 1$, so by (1) $\signt_1 = \signt'_1$. Now suppose that $\signt_i = \signt'_i$ for $1 \leq i \leq \stri$, for some $\stri \geq 1$. If $\stri = |S| = |S'|$ then we are done. Thus assume $\stri < |S|$. We claim that $\signt_{\stri + 1} = \signt'_{\stri + 1}$. Let $\stri' \leq \stri$ be the largest index such that $\signt_{\stri'} \neq 0$; since $\stri' \leq \stri$, we have $\signt'_{\stri'} = \signt_{\stri'}$. Let $\sclen = \sclen(\signt_{\stri'}) = \sclen(\signt'_{\stri'})$; since $S, S' \in \scode^*$, we must have $\signt_i, \signt'_i = 0$ for $i = \stri' + 1, \ldots, \stri' + \sclen$. If $\stri' + \sclen > \stri$ then we are done. Otherwise, by definition of $\stri'$ we must have $\stri' + \sclen = \stri$; but then we must have $\signt_{\stri + 1}, \signt'_{\stri + 1} \geq 1$. Thus by induction $\signt(S) = \signt(S')$. Thus $\signt(S[r]) = \signt(S'[r])$ for $1 \leq r \leq |S| = |S'|$, so that when (2) holds we must have $S[r] = S'[r]$ for $1 \leq r \leq |S|$, i.e., $S = S'$.
\par\vskip\baselineskip\par
\leftline{\bf 2. Application to binary codes}
\par
We give an application of the foregoing results to binary codes. If $b$ is a string on $\{ 0, 1 \}$, we define its parity to be $0$ or $1$ (alternatively, even or odd) as it has an even or odd number of $1$s, respectively. We then have the following result.
\par\vskip0.5\baselineskip\par\noindent
PROPOSITION 3. Let $\calb$ be any collection of bitstrings, let $\calb_0$, $\calb_1$ be any partition of $\calb$ by parity, and suppose that $\calb_0, \calb_1 \neq \emptyset$. Set $\capr_0 = \calb_0$, let $\{ \capr_\capri \}$, $\capri \in \{ 1, \ldots, n \}$, be any partition of $\calb_1$, and let $\csclen : \{ 1, \ldots, n \} \rightarrow \N$. Then there is a solid code $\ccode$ on $\calb$ of codeword length at most $1 + {\rm max}\,\{ \csclen(\calba)\,|\,\calba \in \calb_1 \}$ and size
$$
\sum_{\capri = 1}^n |\capr_\capri| |\calb_0|^{\csclen(\capri)}.
$$
Moreover, if a string $S \in \ccode^*$ is transmitted over a noisy channel which changes at most 1 bit in each bitstring in $\calb$ and the received string $S'$ is also in $\ccode^*$, then $S' = S$.
\par
{\it Proof.} As above, we may take
$$
\ccode = \{ \calba_0 \ldots \calba_k\,|\,\calba_0 \in \capr_\capri,\,\capri \geq 1,\,\calba_i \in \calb_0,\,k = \csclen(\capri) \},
$$
which is seen to be solid by Proposition 1. Since a single bitflip {\it must\/} change the parity of a bitstring, the probability distribution $\cprobcor(\calba \rightarrow \calba')$ induced by the channel noise must vanish when $\calba, \calba' \in \calb_0$ or $\calba, \calba' \in \calb_1$. Now since $\capr_\capri \subset \calb_1$ for $\capri \geq 1$ and $\capr_0 = \calb_0$, this implies that $\cprobcor(\calba\rightarrow\calba') = 0$ if $\signt(\calba) = \signt(\calba')$, so that condition (2) in Proposition 2 holds; similarly, if $\signt(\calba), \signt(\calba') \geq 1$, then $\calba, \calba' \in \calb_1$, so again $\cprobcor(\calba \rightarrow \calba') = 0$ and condition (1) in Proposition 2 holds as well. Thus $S' \in \ccode^*$ implies $S' = S$, as claimed.
\par\vskip0.5\baselineskip\par\noindent
We note that $\ccode$ is only solid when considered as a code {\it on\/} $\calb$, and need {\it not\/} be solid when considered as a code on $\{ 0, 1 \}$.
\par
Informally, Proposition 3 shows that the property of being solid and the requirement that the received bitstring be decodable allow us to rule out errors associated with corruptions $\capr_0 \rightarrow \capr_\capri$, $\capr_\capri \rightarrow \capr_0$ for $\capri \in \{ 1, \ldots, n \}$. This is useful when constructing error-detecting codes on predetermined sets of bitstrings, for example.
\par\vskip\baselineskip\par
\leftline{\bf References}
\par\noindent
\halign to\hsize{\hbox to 2em{#\hfil}\kern 1em&\vtop{\noindent\advance\hsize by -3em #}\hfill\cr
\refWangWang & Wang, Geyang, Wang, Qi. $Q$-ary Non-Overlapping Codes: A Generating Function Approach. {\it IEEE Transactions on Information Theory\/} {\bf 68}, no. 8, 5154-5164 (2022).\cr
\refQinLuo & Qin, Chunyan, Luo, Gaojun. A generalized construction of variable-length non-overlapping codes. {\it Designs, Codes and Cryptography\/} {\bf 93}, 2229-2243 (2025).\cr
\refcodesautomata & Berstel, Jean, Perrin, Dominique, Retenauer, Christophe. {\it Codes and Automata}. Cambridge: Cambridge University Press, 2010.\cr
\refshyryu & Shyr, H.\ J., Yu, S.\ S. Solid codes and disjunctive domains. {\it Semigroup Forum\/} {\bf 41}, 23-37 (1990).\cr
\reflam & Lam, Nguyen Huong. Finite maximal solid codes. {\it Theoretical Computer Science\/} {\bf 262}, 333-347 (2001).\cr
\refhandbook & J\"urgensen, Helmut, Konstantinidis, Stavros. Codes. In {\it Handbook of Formal Languages}, Rosenberg, G.\ (ed.). Berlin: Springer-Verlag, 1997.\cr
\refUnicode & {\it The Unicode Standard, Version 17.0}. South San Francisco: Unicode Consortium, 2025.\cr}
\bye